# Analytical photoresponses of Schottky contact MoS$_2$ phototransistors

Jianyong Wei[1], Yumeng Liu[1], Yizhuo Wang[1], Kai Li[1], Zhentao Lian[1], Maosong Xie[1], Xinhan Yang[2], Seyed Saleh Mousavi Khaleghi[1], Fuxing Dai[3], Weida Hu[3], Xuejiao Gao[2*], Rui Yang[1,4*], Yaping Dan[1*]

[1]State Key Laboratory of Advanced Optical Communication Systems and Networks, University of Michigan – Shanghai Jiao Tong University Joint Institute, Shanghai Jiao Tong University, Shanghai 200240, China

[2]Hebei Key Laboratory of Physics and Energy Technology, Department of Mathematics and Physics, North China Electric Power University, Baoding, Hebei 071003, China

[3]State Key Laboratory of Infrared Physics, Shanghai Institute of Technical Physics, Chinese Academy of Sciences 500 Yu Tian Road, Shanghai 200083, China

[4]State Key Laboratory of Radio Frequency Heterogeneous Integration, Shanghai Jiao Tong University, Shanghai 200240, China

Email: Yaping.dan@sjtu.edu.cn; rui.yang@sjtu.edu.cn; gaoxuejiao@ncepu.edu.cn

Abstract

High-gain photodetectors based on two-dimensional (2D) semiconductors, in particular those in photoconductive mode, have been extensively investigated in the past decade. However, the classical photoconductive theory was derived on two misplaced assumptions. In this work, we established an explicit analytical device model for Schottky contact MoS$_2$ phototransistors that fits well with experimental data. From the fitting results, we found that the Richardson constant of the MoS$_2$ Schottky contact is temperature dependent, indicating that the Schottky contacts for the 2D material is best described by the mixed thermionic emission and diffusion model. Based on this device model, we further established an analytical photoresponse for the few-layer MoS$_2$ phototransistors, from which we found the voltage distribution on the two Schottky contacts and the channel, and extracted the minority carrier recombination lifetimes. The lifetimes are comparable with the values found from transient photoluminescence measurements, which therefore validates our analytical photoresponses for Schottky contact 2D semiconducting phototransistors.

Low-dimensional photodetectors often have extraordinarily high photogain (up to $10^{10}$).[1-5] It was believed in the classical photogain theory that the gain comes from the recycling of photogenerated charge carriers, *i.e.* photogenerated electrons or holes, driven by electric fields, circulate in the circuit multiple times before recombination.[6-10] In fact, photogenerated electrons and holes in a doped semiconducting photoconductor cannot be spatially separated due to the ambipolar transport.[11] In recent years, we found that the classical theory was derived on two misplaced assumptions.[12] The first assumption is that the photogenerated excess carriers are uniform in spatial distribution on condition of uniform doping and illumination, which becomes invalid in the presence of metal-semiconductor boundary confinement. The second assumption is that the excess electrons and holes equally contribute to the photocurrent, which intuitively should be true because the absorption of photons will create an equal number of electrons and holes. However, devices in practice often have surfaces or interfaces where depletion regions and/or trap states appear. The depletion regions and/or trap states will localize excess minority carriers, leaving the majority counterparts to dominantly contribute to the photocurrent.

After correcting these two assumptions, we further derived the explicit gain equations for single crystalline nanowires based on photo Hall measurements.[13-14] The derived gain equations are a function of light intensity and device physical parameters such as doping concentration, nanowire diameter and surface depletion width. The gain equations fit well the experimental data from which we extracted parameters including minority carrier recombination lifetimes that are consistent with experimental results in literature.[15-16] Although the photo gain is still proportional to the ratio of minority recombination lifetime to transit time, the explicit gain equations show that the high photo gain ($10^6$ - $10^8$) does not originate from this ratio since the ratio is not more than 10. Instead, the high photo gain comes from the light-illumination-induced photovoltage across the surface depletion region that modulates the conduction channel width.

High-gain photodetectors based on two-dimensional (2D) semiconductors have been extensively investigated.[3-4, 17-22] However, there is no analytical theory that governs the photoresponses of these devices, and the effects from the Schottky contact in 2D phototransistors have been largely ignored in the past decades. Inspired by our previous work on high-gain photoconductors, we found in this work that high-gain photoresponses are created by the light-induced modulation of Schottky barrier height in $MoS_2$ transistors. Analytical photoresponses were established on a modified Schottky junction model for two-dimensional (2D) semiconductors. The experimental data are well fitted with the analytical photoresponse model, from which we extracted the minority recombination lifetimes. Transient photoluminescence was employed to independently find the minority recombination lifetimes that are largely comparable with the extracted values from our analytical photoresponse theory, which further validates the model. Based on the model, we analyzed the voltage distribution among the two contacts and the channel region. The results are important for understanding the carrier dynamics towards high-performance 2D photodetectors, and for precisely modeling the photoresponses of 2D phototransistors.

**RESULTS AND DISCUSSION**

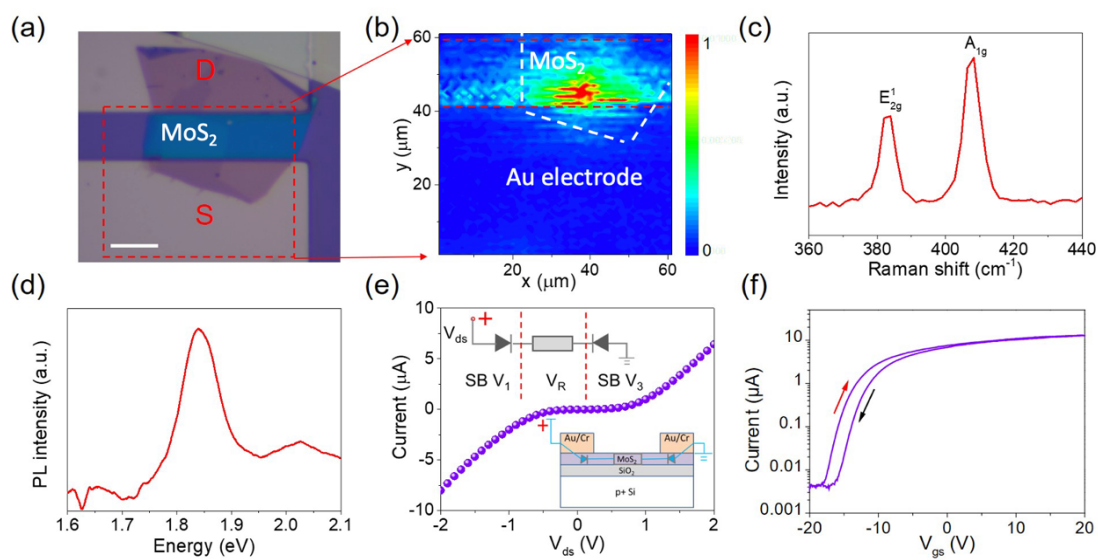

Figure 1. (a) Optical microscopic image of a $MoS_2$ device. Scale bar is 20 μm. (b) Scanning photocurrent microscopy of the $MoS_2$ device. (c) Raman shifts and (d) photoluminescence spectrum of the multilayer $MoS_2$. (e) $I$-$V$ characteristics of the $MoS_2$ device. Inset: device model of the Schottky contact $MoS_2$ (top inset) and schematic of the device structure (bottom inset). (f) Gate transfer characteristics of the $MoS_2$ transistor with $V_{ds}$ = 2 V.

The optical image of the back-gate $MoS_2$ field effect transistor (FET) is shown in Fig. 1a, from which we can see that the multilayer $MoS_2$ is on top of and in good contact with both electrodes. The device was fabricated following the procedure as described below. Two Au/Cr electrodes were first formed on $p^+$-Si/$SiO_2$ wafer (300nm thick $SiO_2$) by photolithography and thermal evaporation.

As the next step, a multilayer MoS₂ flake was mechanically exfoliated from a bulk MoS₂ and then dry-transferred onto the target surface in contact with the electrodes. Finally, the device was annealed in vacuum at 300°C for 30 minutes. The channel is ~ 20 μm long and ~ 64 μm wide. Fig.1b shows the scanning photocurrent microscopy of the device, which we will discuss later.

The material quality of the exfoliated MoS₂ flakes were examined with Raman and photoluminescence spectroscopy. As shown in Fig. 1c, two Raman peaks at 383.9 and 408.3 cm$^{-1}$ were observed, which are the $E^1_{2g}$ and $A_{1g}$ vibration mode of MoS₂, respectively. It is known that the separation between $E^1_{2g}$ and $A_{1g}$ mode evolves with the number of MoS₂ layers. A peak separation of 24.4 cm$^{-1}$ indicates that the flake is a few-layer MoS₂.[22] Photoluminescence (PL) characterizations were carried out under the excitation of a 532 nm laser. The single strong PL peak at 1.84 eV was attained as shown in Fig. 1d, indicating a good crystalline quality of the MoS₂ flake.

Schottky contacts often form between 2D materials and metal electrodes, mainly due to the Fermi level pinning at the interface and the mismatch in Fermi energy levels of the 2D material and metal. It is well known that Schottky junctions play a dominant role in the electronic properties of semiconductor devices.[22-28] As a result, it is not surprising to see that the current-voltage characteristics of our device are nonlinear (Fig.1e). To double check the observation, we performed the scanning photocurrent microscopy (SPCM) on the device. In this measurement, a focused laser beam (λ=532nm) was used to scan over the MoS₂ device. Photocurrents and laser positions were then recorded simultaneously, forming a raster photocurrent map. The photocurrent map in Fig.1b (placed right next to the device image in Fig.1a for comparison) shows that the photocurrent of the MoS₂ device at a bias of 2V is mainly located near the bottom Au electrode and extends well into the channel. This is because the Schottky junction near the bottom electrode is reverse biased and the top Schottky junction is at forward bias. At a bias of 2V, the resistance of the forward Schottky junction is negligible. The reverse biased Schottky junction and channel are the main components in device resistance (see more discussions later) and therefore are the main sources of photocurrent in the map.

Fig.1f shows the gate-transfer characteristics of the MoS₂ device at a source-drain bias of 2V. At this bias, the MoS₂ channel dominates over the device resistance. As the gate voltage sweeps from 20V to -20V, the source-drain current drops by 3 orders of magnitude, indicating that the MoS₂ channel is n-type. The hysteresis in gate transfer characteristics is attributed to the defects that trap or emit charges.

The *I-V* curve of a Schottky junction is governed by eq. (1):

$$I_0 = I_{s0} exp\left(-\frac{\phi}{kT}\right)\left[e^{\frac{qV}{kT}} - 1\right] \quad (1)$$

where $\phi$ is the Schottky barrier height (SBH), $k$ the Boltzmann constant, T the temperature and $V$ the applied voltage. $I_{s0}$ is the current constant given by $I_{s0} = A^*T^{3/2}S$ for 2D semiconductors with $A^*$ being the Richardson constant, $S$ the cross-sectional width of MoS$_2$ channel.

Schottky junctions of 2D materials in contact with metal, although fabricated in the same process, often have different *I-V* characteristics due to the dielectric disorder of 2D materials[29], which is reflected in the difference of ideality factor and barrier height. More importantly, the Schottky barrier height of the 2D semiconductor can be modulated by the bias applied to the barrier following eq.(2).[30]

$$\phi(V) = \phi_0 \pm qV\left(1 - \frac{1}{n}\right) \quad (2)$$

where $\phi_0$ is the intrinsic barrier height at zero bias, $V$ the applied bias, $n$ the ideal factor and $\pm$ for forward and reverse bias, respectively. When we plug eq.(2) into eq.(1), we can rewrite the forward current as eq.(3a) and the reverse current as eq.(3b) after the term $e^{\frac{qV}{kT}} - 1$ is simplified as $e^{\frac{qV}{kT}}$ for forward bias and -1 for reverse bias, respectively, when |V|>3kT/q.

$$I_F = I_s exp\left(\frac{qV_F}{nkT}\right) \quad (3a)$$

$$I_R = -I_s exp\left[\frac{qV_R}{kT}\left(1 - \frac{1}{n}\right)\right] \quad (3b)$$

in which $I_s = I_{s0} exp\left(-\frac{\phi_0}{kT}\right)$, $V_F$ is the forward bias and $V_R$ the reverse bias.

Our device consists of two back-to-back Schottky junctions (#1 and #2) and one channel resistor $R$ (as shown in the inset of Fig.1e). A bias voltage $V$ will create a current $I$ flowing through the forward junction (#1), channel resistor and the reverse junction (#2). The bias voltage $V$ will distribute among these three components in series, i.e. $V = V_{F1} + V_{R2} + R \times I$, which can be written as eq.(4) after plugging $V_F$ for Junction #1 and $V_R$ for Junction #2 found from eq.(3) (Note $I = I_F = -I_R$):

$$V = n_1 \frac{kT}{q} ln\left(\frac{I}{I_{s1}}\right) + \frac{n_2}{n_2-1}\frac{kT}{q} ln\left(\frac{I}{I_{s2}}\right) + RI \quad (4)$$

in which the subscript 1 and 2 represents the corresponding parameter of Junction #1 and #2,

respectively. More specifically, $I_{s1} = I_{s01} exp\left(-\frac{\phi_{01}}{kT}\right)$ and $I_{s2} = I_{s02} exp\left(-\frac{\phi_{02}}{kT}\right)$. To make it convenient for fitting with experimental data, eq.(4) is reformatted into eq.(5) in which we use the current *I* as the variable and the bias voltage *V* as the function.

$$V = \left(n_1 + \frac{n_2}{n_2-1}\right)\frac{kT}{q}\ln\frac{I}{I_s^e} + RI \qquad (5)$$

, where $I_s^e = I_{s1}^{\alpha} I_{s2}^{\beta}$ with $\alpha = n_1/(n_1 + \frac{n_2}{n_2-1})$ and $\beta = \frac{n_2}{n_2-1}/(n_1 + \frac{n_2}{n_2-1})$. The derived eq.(5) governs *I-V* characteristics of back-to-back $MoS_2$ Schottky junction, which is universal for other 2D semiconductor devices in Schottky contact.

We plot the forward and reverse voltage *vs* current (*V-I*) characteristics at different gate voltage. Excitingly, the *V-I* characteristics in Fig.2a can be nicely fitted with eq.(5), from which we extracted the leakage current $I_S$ (calculated from the extracted $I_S^e$, see Table S1 in SI Section I) for the two Schottky junctions and the channel resistance *R*. Fig. 2b shows that the extracted leakage current $I_{s1}$ and $I_{s2}$ for Junction #1 and #2, respectively, which have a competing response to the gate voltage. Fig. 2c plots the extracted channel resistances as a function of gate voltage. As expected, the channel resistance *R* extracted from the forward and reverse bias are comparable, which validates our device model. What's more, the extracted ideality factor $n_1$ and $n_2$ are both slightly larger than 1 (see Fig. S1a in SI Section I), which is reasonable because of the thermally-assisted tunneling in Schottky junctions.

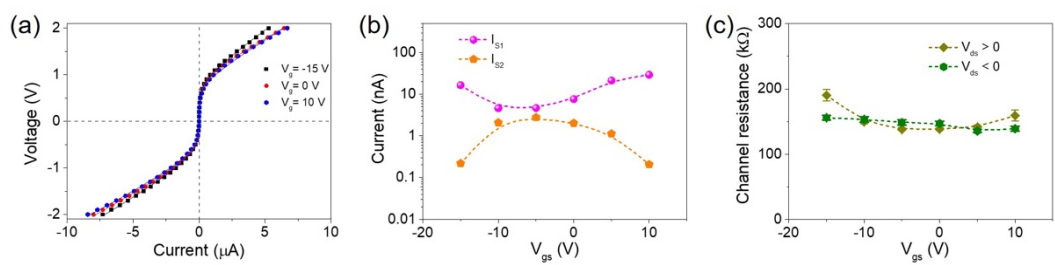

Figure 2. (a) Gate-dependent *V-I* characteristics fitted with eq.(5). (b) Extracted the leakage current $I_{s1}$ (purple diamonds) and $I_{s2}$ (Orange dots) for the two Schottky junctions. (c) Extracted channel resistances for source-drain bias greater (brown diamonds) and less than zero (green dots).

Figure 3 shows the temperature dependent *V-I* characteristics for forward and reverse bias which can be also nicely fitted with eq.(5). From the fittings, we extracted the leakage currents and channel resistances as a function of temperature, as shown in Fig. 3b and c, respectively. The leakage currents for both junctions increase as the temperature rises, but with different correlations. It is worthy to point out that we cannot extract the intrinsic Schottky barrier height $\phi_0$ from the leakage currents $I_{s1}$ or $I_{s2}$ because the Richardson constant in $I_{s0}$ is temperature dependent (see Fig.S2 in SI Section II). This fact indicates that the metal-MoS$_2$ Schottky junction shall be modeled with the mixed thermionic emission and diffusion model developed by Crowell and Sze,[31] in which the Richardson constant is implicitly dependent on temperature. Moreover, the temperature dependent Richardson constant also explains why the channel resistances $R$ calculated with $V_{ds} > 0$ and $V_{ds} < 0$ deviate from each other as the temperature reduces to 175K or lower.

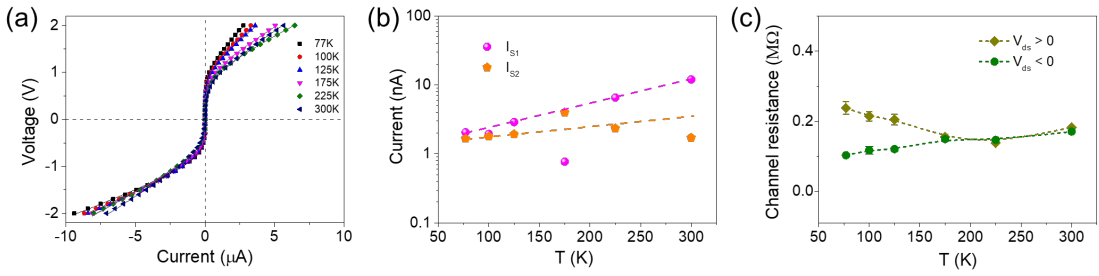

Figure 3. (a) Temperature-dependent *V-I* characteristics fitted with eq.(5). (b) Extracted the leakage current Is$_1$ (purple diamonds) and Is$_2$ (Orange dots) for the two Schottky junctions. (c) Extracted channel resistances for source-drain bias greater (brown diamonds) and less than zero (green dots).

Based on the device theory above, we investigated the working mechanism and then photoresponses of the MoS$_2$ phototransistor. First, we recorded the *I-V* characteristics (a different device) under illumination of light (λ = 525 nm) with different intensities (Fig. 4a). All the six *I-V* curves with light intensity increasing from 0 to 24 mW/cm$^2$ are well fitted with eq.(5). The derived voltage load on Junction #1 ($V_1$), Junction #2 ($V_3$) and channel resistance ($V_R$) as a function of source-drain bias are shown in Fig.4b. At bias voltage higher than 1.8 V, the channel resistance dominates the source-drain resistance. But as the bias is lower than 1.8 V, the reverse biased Schottky junction (here junction #2) play a leading role. Under light illumination, the photoresponse of reverse biased Schottky junction keeps predominant. This photoresponse is ascribed to the reduction $\Delta\phi$ of the intrinsic Schottky barrier height $\phi_{02}$. Therefore, we have the total current under light illumination shown in eq.(6).

$$I_L = -I_{s0} \exp\left(-\frac{\phi_{02}-\Delta\phi}{kT}\right) \exp\left[\frac{qV_R}{kT}\left(1-\frac{1}{n}\right)\right] = I_{dark} \exp\left(\frac{\Delta\phi}{kT}\right) \quad (6)$$

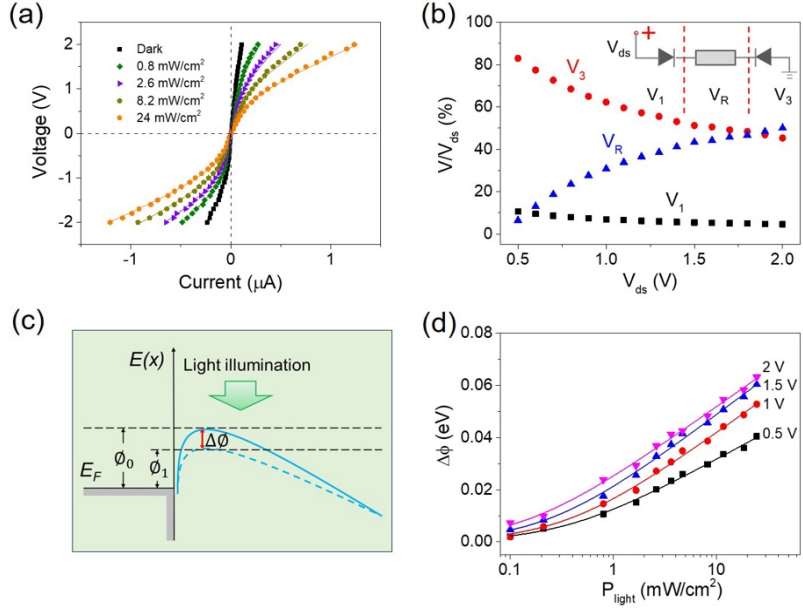

Figure 4. (a) *V-I* characteristics under light illumination (λ = 525 nm) of different intensity. (b) Voltage drops on the forward and reverse biased Schottky Junction, and the MoS$_2$ channel in darkness. (c) Schottky barrier reduction as a function of light illumination for different bias.

Previously, we found that the nanowire channel will be pinched off when the nanowire is narrower than two times of the surface depletion region width in the nanowire. In this case, a potential barrier exists between source and drain. Light illumination will reduce the potential barrier, inducing a photocurrent. The light intensity and the potential barrier reduction is correlated following eq.(7).[14, 32] Schottky contact MoS$_2$ phototransistor in this work is quite similar, with a potential barrier from the predominant reverse biased Schottky junction. At a fixed bias, light illumination will also reduce this potential barrier (see Fig.4c). The potential barrier reduction $\Delta\phi$ can be calculated from the current in darkness and under light illumination following eq.(6). Fig.4d shows $\Delta\phi$ as a function of light intensity at different bias which can be nicely fitted with eq.(7). The fitting results are consistent with what will be presented later in Fig.5.

$$P_{light} = P_{light}^{S}\left[\exp\left(\frac{\Delta\phi}{mkT}\right) - 1\right] \quad (7)$$

, where *m* is the ideality factor determining how effectively the potential barrier is reduced by light illumination. It depends on the device structure, light absorption, defects in the junction and others. $P_{light}^{S}$ is defined as the critical light intensity which is given by $P_{light}^{S} = \frac{\hbar\omega n_i}{2\alpha\tau_0}$, in which $\hbar\omega$ is the photon energy, $n_i$ is the intrinsic electron concentration (per unit area) of MoS$_2$, $\alpha$ is the light absorption ratio by MoS$_2$ and $\tau_0$ is the minority recombination lifetime in MoS$_2$.

From eq. (6) and (7), we find that the total current under light illumination is dependent on the illumination light intensity $P_{light}$ following eq.(8).

$$I_L = I_{ph} + I_{dark} = I_{dark}\left(\frac{P_{light}}{P_{light}^S} + 1\right)^m \quad (8)$$

Furthermore, the photogain (G) shown in eq.(9) can be derived from eq.(8).

$$G = \frac{I_{ph}/q}{P_{light}WL/\hbar\omega} = G_{max}\left\{\frac{P_{light}^S}{mP_{light}}\left[\left(\frac{P_{light}}{P_{light}^S} + 1\right)^m - 1\right]\right\} \quad (9)$$

, where $G_{max} = \frac{m\hbar\omega}{qP_{light}^S WL}$ is the maximum gain when the light intensity approaches to zero, $W$ and $L$ are the width and length of the MoS$_2$ channel.

Fig.5a along with the inset exhibits the experimental current of the MoS$_2$ device under light illumination and the corresponding photogain as a function of light intensity at different bias, which can be well fitted with eq.(8) and (9), respectively. In eq.(9), the photocurrent is normalized against the light intensity to derive the photogain. As a result, the photogain maximizes at weak light intensity. At weak light intensity, the signal-to-noise ratio is small for experimental data. A large uncertainty will be created if we use eq.(9) to fit the experimental data. For this reason, we use Eq.(8) to perform our fittings (solid curves). At high bias (> 1.8 V in our case), although the channel resistance starts to dominate and the reverse biased Schottky junction become less important, the theory still fits well the experimental data. This indicates that the photoresponse of the MoS$_2$ channel also follows eq.(8). This is likely reasonable considering that there are defects in the atomically thin MoS$_2$ flake. These defects are charged, forming a potential barrier reducing the effective conduction channel width. Light illumination will reduce the potential barrier and create a photocurrent, following similar equations as shown in eq.(7) and (8).[33] A systematic investigation of the analytical photoresponses in the channel is subject to future work.

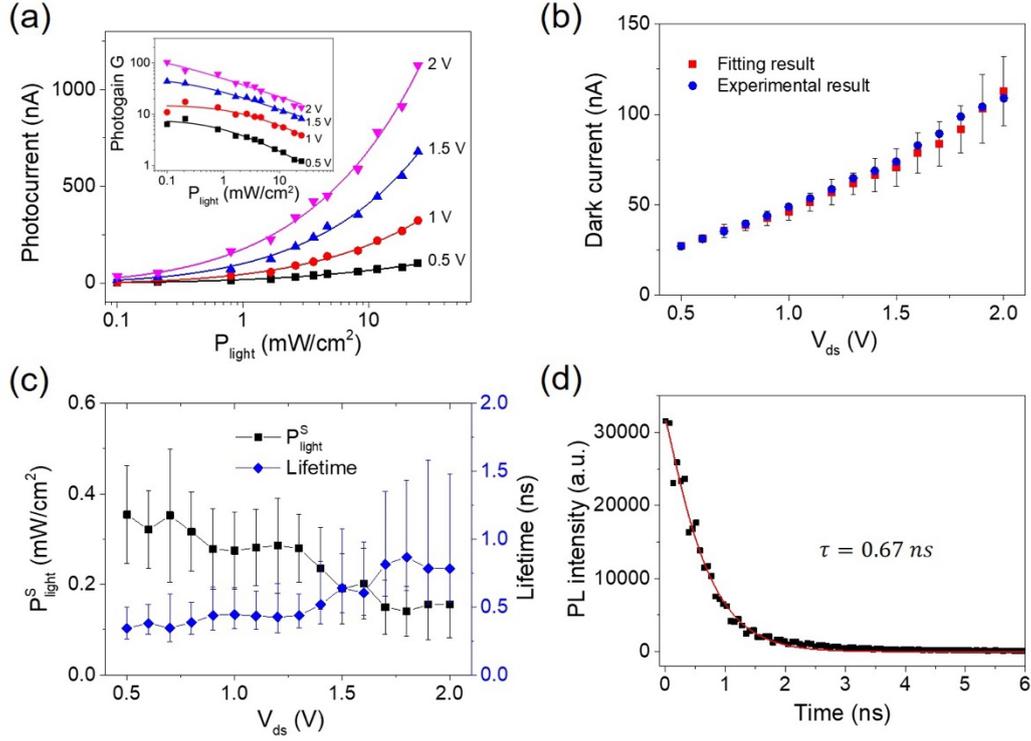

Figure 5 (a) Photocurrent dependent on light intensity at different source-drain bias. Inset: photogain as a function of light intensity at different source-drain bias. (b) Experimental and extracted dark current vs source-drain bias. (c) The extracted critical light intensity $P_{light}^S$ and minority carrier lifetime $\tau_0$ vs source-drain bias. (d) Transient photoluminescence of the MoS$_2$ flake.

Here, let us focus on the fitting results of eq.(8). From the fittings, we extract the dark current $I_{dark}$, the ideality factor $m$ and the critical light intensity $P_{light}^S$. The extracted dark currents are almost identical with the experimental value (Fig.5b), which validates the analytical photoresponse model. The extracted ideality factor $m$ remains largely constant around 0.35 to 0.48. To calculate the minority carrier recombination lifetime $\tau_0 = \frac{\hbar \omega n_i}{2\alpha P_{light}^S}$, we first find the photon energy of $\hbar \omega = 2.36 eV$ for the incident light at a wavelength of 525nm, and the absorption ratio of approximately $\alpha = 41\%$ for a multilayered MoS$_2$ flake according to previous work.[33] The intrinsic electron concentration $n_i$ of MoS$_2$ depends on the effective mass of electrons and holes. It is best estimated around $2.65 \times 10^5$ cm$^{-2}$ (see SI section III). Based on the extracted value and the expression of the critical light intensity $P_{light}^S$ given above, we calculate the minority recombination lifetime of MoS$_2$ in a range of 0.3 ns to 1 ns as shown in Fig.5c. The minority recombination lifetime increases as the source-drain bias likely because the photoresponse of the MoS$_2$ channel plays a more important role at higher voltage. Note that a metal in contact with a semiconductor will bring additional defects. Therefore, it is expected that the minority recombination lifetime is longer in the channel.

To further validate the analytical photoresponse, we experimentally measured the minority recombination lifetime. A pulsed laser illumination ($\lambda$ = 405 nm) was employed to excite excess carriers in the MoS$_2$ flake. When the excitation pulse is switched to off state, the photoluminescence will exponentially decay following a stretched exponential decay function (Fig.5d).[34] Fitting the transient decay with this stretched exponential function, we find the minority carrier recombination lifetime as 0.67 ns, which is comparable with the lifetimes we extracted from the analytical photoresponses (Fig.5c), in particular when the MoS$_2$ channel resistance starts to dominate at high bias (> 1.8 V).

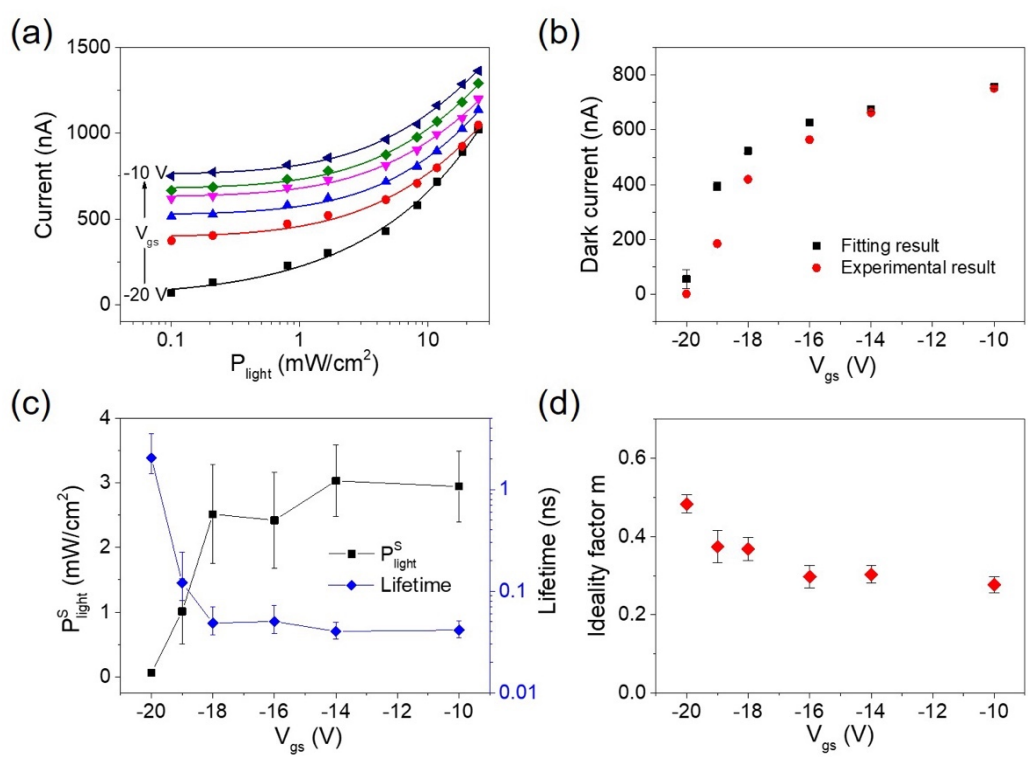

Figure 6 (a) Total current under light illumination dependent on light intensity for different gate voltages. (b) Extracted dark current (black squares) in comparison with experimental ones (red dots) at different gate voltage. (c) Extracted critical light intensity and minority recombination lifetime at different gate voltage. (d) Extracted ideality factor *m* at different gate voltages.

The gate voltage also has a significant impact on the photoresponse of the MoS$_2$ device. We recorded the total current under light illumination for different gate voltage as shown in Fig.6a (a different device). The current dependence on the light intensity can be nicely fitted with eq.(8) for all gate voltages. The extracted dark currents (black squares) are almost the same with the experimental ones (red dots) as shown in Fig.6b, suggesting that eq.(8) is valid for gated $MoS_2$ devices. From the extracted critical light intensity (black squares in Fig.6c), we found that the minority carrier recombination lifetime increases from ~ 40 ps to ~ 1 ns (blue diamonds) when the gate voltage sweeps from -10 V to -20V. The initial lifetimes are smaller by an order of magnitude than the ones presented in Fig.5 probably because this new $MoS_2$ flake is thinner or has a poorer quality than the previous one. At high negative gate voltage (< - 18V), the minority recombination lifetime quickly increases likely due to the reduction of majority electrons in $MoS_2$ that suppresses the recombination rate of excess minority holes. Fig.6d plots the ideality factor $m$ at different gate voltages. As the gate voltage sweeps from -10 V to -20 V, $m$ first remains nearly constant around 0.3 and then increases to 0.5 after -18 V, exhibiting a pattern similar to the minority recombination lifetime in Fig.6c.

**CONCLUSION**

In this work, we established an analytical device principle for Schottky contact $MoS_2$ field effect transistors that fits well with experimental *I-V* characteristics. From the fitting results, we found that the Richardson constant of the $MoS_2$ Schottky contact is temperature dependent, indicating that the Schottky contacts for the 2D material is best described by the mixed thermal emission and diffusion model. After that, we also established an analytical photoresponse equation for the $MoS_2$ phototransistors, from which we extracted the minority recombination lifetimes, which are consistent with the transient photoluminescence measurements. These established principles and equations are not unique to $MoS_2$ device and may be universally applicable to other transistors based on 2D semiconductors.

**EXPERIMENTAL SECTION**

**Device Fabrication** $MoS_2$ devices were fabricated on a highly doped p-type Si wafer with 300 nm thick $SiO_2$ on the top. As the first step, the wafer was cleaned with acetone and deionized (DI) water. Next, the source and drain electrodes were formed on the substrate by photolithography and thermal evaporation. For the photolithography process, NR9-1500PY (Futurrex Inc. USA) photoresist was first spin-coated on the substrate at 4000 rpm for 40 s. After baked at 140 °C for 1 min, the NR9 photoresist was exposed to UV light (MDA-400) for 14 s and developed in the developer after post-baking at 110 °C for 1 min. Then the sample was transferred in an evaporator (Thermal Evaporator, Angstrom Engineering) and 5/50 nm thick Cr/Au was evaporated on the sample in high vacuum. A liftoff process was conducted in the acetone to create an array of Cr/Au electrodes. In the following step, a multilayer $MoS_2$ flake was mechanically exfoliated from a bulk $MoS_2$ piece and then transferred onto the target surface. Finally, the device was annealed in vacuum at 300°C for 30 minutes.

**Optoelectronic Measurements** The devices were characterized in a vacuum probe station by high-precision digital sourcemeters (Keithley 2400 and 2636). A 525 nm commercial LED was used as the light source. The light intensity is controlled by the driving current of the circuit. The light intensity is calibrated by a commercial photodiode (G10899-003 K, Hamamatsu).

ACKNOWLEDGEMENT

This work was financially supported by the Oceanic Interdisciplinary Program of Shanghai Jiao Tong University (No. SL2022ZD107), the Shanghai Jiao Tong University Scientific and Technological Innovation Funds (No. 2020QY05), Science and Technology Commission of Shanghai Municipality (STCSM) Rising-Star Program (Grant 23QA1405300), Lingang Laboratory Open Research Fund (Grant LG-QS-202202-11), and the Shanghai Pujiang Program (No. 22PJ1408200). The devices were fabricated at the Center for Advanced Electronic Materials and Devices (AEMD), Shanghai Jiao Tong University.

AUTHOR DECLARATIONS

The authors have no conflicts to disclose.